\documentclass[pra,aps,amsmath,showpacs,twocolumn,floatfix,
  superscriptaddress]{revtex4}
\usepackage{bm}
\usepackage{amsmath}
\usepackage{amssymb}
\usepackage{epsfig}
\usepackage{dcolumn}
\newcommand{\kappaprime}{\kappa^\prime}
\newcommand{\kperp}{k_{\!_\perp}}
\newcommand{\kperpprime}{k_{\!_\perp}^\prime}
\newcommand{\kz}{k_z}
\newcommand{\kzprime}{k_z^\prime}

\newcommand{\rperp}{r_{\!\!_\perp}}
\newcommand{\Rperp}{R_{\!_\perp}}
\newcommand{\rperpprime}{r_{\!_\perp}^\prime}
\newcommand{\rperpmin}{r_{\!_\perp\mathit{min}}}
\newcommand{\rperpmax}{r_{\!_\perp\mathit{max}}}

\newcommand{\varphiprime}{\varphi^\prime}
\newcommand{\zprime}{z^\prime}
\newcommand{\muprime}{{\mu^\prime}}
\newcommand{\ekin}{e_{\mathit{kin}}}
\newcommand{\erot}{e_{\mathit{rot}}}
\newcommand{\eself}{e_{\mathit{self}}}
\newcommand{\etot}{e_{\mathit{tot}}}
\newcommand{\eHO}{e_{\mathit{HO}}}
\newcommand{\aHO}{a_{\mathit{HO}}}
\newcommand{\Vext}{V_{\mathit{ext}}}
\begin{document}
\title{Thomas-Fermi approximation to static vortex states in superfluid
  trapped atomic gases}
\author{Michael Urban}
\affiliation{Institut de Physique Nucl{\'e}aire, F-91406 Orsay
  Cedex, France}
\author{Peter Schuck}
\affiliation{Institut de Physique Nucl{\'e}aire, F-91406 Orsay
  Cedex, France}
\affiliation{Laboratoire de Physique et Mod\'elisation des Milieux
  Condens\'es, CNRS \& Universit\'e Joseph Fourier, Maison des Magist\`eres,
  B.P. 166, F-38042 Grenoble Cedex 9, France}
\author{Xavier Vi{\~n}as}
\affiliation{Departament d'Estructura i Constituents de la
  Mat{\`e}ria, Facultat de F{\'{\i}}sica, Universitat de Barcelona, Diagonal
  647, E-08028 Barcelona, Spain}
\begin{abstract}
We revise the Thomas-Fermi approximation for describing vortex states
in Bose condensates of magnetically trapped atoms. Our approach is
based on considering the $\hbar\to 0$ limit rather than the
$N\to\infty$ limit as Thomas-Fermi approximation in close analogy with
the Fermi systems. Even for relatively small numbers of trapped
particles we find good agreement between Gross-Pitaevskii and
Thomas-Fermi calculations for the different contributions to the total
energy of the atoms in the condensate. We also discuss the application
of our approach to the description of vortex states in superfluid
fermionic systems in the Ginzburg-Landau regime.
\end{abstract}
\maketitle
%
%%%%%%%%%%%%%%%%%%%%%%%%%%%%%%%%%%%%%%%%%%%%%%%%%%%%%%%%%%%%%%%%%%%%%%
\section{Introduction}
The discovery of Bose-Einstein condensation in trapped alkali-metal
gases at ultra-low temperature \cite{Anderson,Davis,Bradley} has
developed a huge amount of experimental and theoretical
investigations. The experimental conditions are such that the atomic
gas is at very low density and that the interactions can be
parametrized in terms of a scattering length $a$. In this situation a
mean-field description through the Gross-Pitaevskii equation (GPE)
\cite{Pitaevskii,Gross} is able to give, at least at low temperature,
a precise description of the atomic condensates and their dynamics
\cite{Dalfovo96a,Dalfovo96b,Parkins,Dalfovo99}.

One important question concerns the superfluid character of the Bose
condensates. Among other properties, the existence of quantum vortices
is a signal of the superfluidity. The possibility of trapped quantized
vortices was one of the primary motivations of the GP theory
\cite{Pitaevskii,Gross} and some amount of theoretical work about this
topic has been developed during the last years
\cite{Dalfovo96a,Dalfovo96b,Dalfovo99,Fetter,Rokhsar,Guilleumas}. The
experimental evidence of such quantized vortices has recently been
verified \cite{Matthews,Madison}.

Since the number $N$ of atoms involved in the condensate is generally
large, it is natural to think that the Thomas-Fermi (TF) approach can
be applied extensively in some aspects of the Bose-Einstein
condensation in traps. This TF limit is usually identified with the
limit of number of atoms $N$ going to infinity rather than to be
interpreted as the $\hbar\to 0$ limit as it happens in the case of
Fermi statistics. Recently, the TF approximation for the ground state
of Bose-Einstein condensates of magnetically trapped atoms has been
discussed as the $\hbar\to 0$ limit \cite{SchuckVinas}. From this
point of view the TF kinetic energy, which is dropped in the
$N\to\infty$ limit of the ground-state calculation, can be obtained
for any number of particles. In this $\hbar\to 0$ limit, a good
agreement between the GP and TF kinetic energies is found even for low
and intermediate number of particles. With the interpretation of the
TF approach as the $\hbar\to 0$ limit, it is also possible to perform
semiclassical TF calculations for the ground state of Bose-Einstein
condensates of atoms with negative scattering length ($^7$Li atoms)
and to compute the excitation energy of collective monopole and
quadrupole oscillations where the kinetic energy of the ground state
of the condensate plays a crucial role \cite{SchuckVinas}.

The TF limit considered as $N\to\infty$ limit has also been applied to
the description of vortex states \cite{Fetter,Rokhsar,Ho,Sinha}. In
this case it is assumed that the radial and axial kinetic energies can
be neglected, and only the rotational kinetic energy is retained
\cite{Ho}. This approximation, however, gives a bad description of the
vortex-core region. A better description of this region in the limit
of large $N$ can be achieved by splitting the condensate wave function
into a product of a slowly-varying envelope, which is obtained by
completely neglecting the kinetic energy, times the solution of the
GPE describing a vortex in homogenous matter \cite{Modugno}. In
contrast to these large-$N$ methods, in this paper, we will again
consider the TF approximation as the $\hbar\to 0$ limit in order to
describe the vortex state semiclassically. In addition to the formal
aspects, this approach has the practical advantages mentioned above,
i.e., that one can calculate the kinetic energy, that it can therefore
be used also in the attractive case, and that it works well also in
the case of relatively few particles.

The experimental and theoretical achievements in Bose-Einstein
condensation have also triggered the investigation of trapped Fermi
gases at very low temperatures \cite{DeMarco,Truscott,Schreck}. One of
the most important goals of the experiments is to reach the BCS
transition to the superfluid phase, associated with the appearance of
a macroscopic order parameter of strongly correlated Cooper pairs in
dilute gases of trapped fermionic atoms. Several theoretical studies
about this topic have recently been developed
\cite{Houbiers,Farine}. In the case where the critical temperature is
much higher than the spacing between the levels in the trap, the
macroscopic order parameter can be obtained through the
Ginzburg-Landau equation (GLE) \cite{Baranov}, which is formally
equivalent to the GPE. In a recent publication \cite{Rodriguez} also
vortex states were discussed within the framework of the GLE. As a
second application of our TF approach we will briefly discuss vortex
states in a superfluid gas of trapped fermionic atoms. Due to the
analogy between the GPE and the GLE our semiclassical approach can
immediately be transferred to this problem.

The paper is organized as follows: In the second section we establish
the TF theory projected on states of defined $z$ component of the
angular momentum and apply it to describe vortex states of a
non-interacting Bose condensate. In the third section we include the
interaction among the atoms in the trap and compare our semiclassical
prediction with the results obtained from the quantal solution of the
GPE for several typical examples. The fourth section is devoted to the
discussion of vortex states in superfluid trapped Fermi systems. Our
conclusions are laid out in the last section.
%
%%%%%%%%%%%%%%%%%%%%%%%%%%%%%%%%%%%%%%%%%%%%%%%%%%%%%%%%%%%%%%%%%%%%%%%%%%%%
\section{The Thomas-Fermi approximation to static vortex states}
We start by considering states having a vortex line along the $z$ axis
and all the atoms flowing around it with quantized circulation. The
order parameter can be written in the form \cite{Dalfovo96a,Rokhsar}
\begin{equation}
\Phi(\vec{r}) = \phi(\rperp,z)\, e^{i\kappa\varphi}\,,
\label{formvortex}
\end{equation}
where $\rperp$ and $z$ are the radial and axial coordinates, $\varphi$
is the angle around the $z$-axis, $\kappa$ is an integer, and
$\phi(\rperp,z) = \sqrt{\rho(\rperp,z)}$, $\rho(\rperp,z)$ being the
density. The vortex state has a tangential velocity $v = \hbar
\kappa/(m \rperp)$ where $\kappa$ is the quantum of circulation, and
the angular momentum along the $z$ axis is $N\hbar \kappa$. The
function $\phi(\rperp,z)$ is obtained as the solution of the following
non-linear Schr{\"o}dinger equation
\begin{multline}
\Big[
  -\frac{\hbar^2}{2m}\Big(
  \frac{\partial^2}{\partial\rperp^2}+
  \frac{1}{\rperp}\,\frac{\partial}{\partial\rperp}+
  \frac{\partial^2}{\partial z^2}\Big)+
  \frac{\hbar^2 \kappa^2}{2 m \rperp^2}
\\
 +\Vext(\rperp,z)+
  g\, \phi^2(\rperp,z)
  \Big]\phi(\rperp,z) = \mu\, \phi(\rperp,z)\,,
\label{gpe}
\end{multline}
which is the GPE for the static vortex state problem. In Eq.
(\ref{gpe}), $\Vext$ is an external potential which for simplicity we
have considered to be a spherical harmonic oscillator (HO) with
frequency $\omega$,
\begin{equation}
\Vext(\vec{r}) = \frac{1}{2}m\omega^2 (\rperp^2+z^2)\,.
\end{equation}
The coupling constant is given by $g = 4\pi \hbar^2 a/m$ with $m$ the
atomic mass and $a$ the $s$-wave scattering length.

For the remaining part of this section, we will concentrate on the
non-interacting case, i.e., $V(\vec{r}) =
\Vext(\vec{r})$. The effect of interactions will be
considered in the next section. For non-interacting particles one
recovers the case of a stationary Schr{\"o}dinger equation for the
harmonic oscillator potential, which is solved by
\begin{equation}
\Phi_\kappa(\vec{r}) = \sqrt{\frac{N}{\pi^{3/2}\,\kappa!\,\aHO^3}}\,
  \Big(\frac{\rperp}{\aHO}\Big)^\kappa\, e^{-(\rperp^2+z^2)/(2\aHO^2)}\,
  e^{i\kappa\varphi}
\label{solutionho}
\end{equation}
with the HO length $\aHO$ defined by
$\aHO = \sqrt{\hbar/(m\omega)}$. The corresponding energy
eigenvalue is given by
\begin{equation}
\mu = \Big(\frac{3}{2}+\kappa\Big)\hbar \omega\,.
\label{muho}
\end{equation}

To derive the TF approximation to the quantal solution of the
non-interacting vortex state (\ref{solutionho}), we start from the
complete set of eigenfunctions of $\hat{\vec{p}}^2$, $\hat{p}_z$ and
$\hat{L}_z$
\begin{equation}
\langle\vec{r}|\kz,\kperp,\kappa\rangle = J_\kappa(\kperp\rperp)\,
  e^{i\kappa\varphi}\, e^{i\kz z},
\end{equation}
which are normalized to
\begin{multline}
\int\! d^3 r\,\langle\vec{r}|\kz,\kperp,\kappa\rangle\,
  \langle \kzprime,\kperpprime,\kappaprime|\vec{r}\rangle\\
  = \frac{4\pi^2}{\kperp}\, \delta(\kperp-\kperpprime)\,
    \delta(\kz-\kzprime)\, \delta_{\kappa\kappaprime}\,.
\end{multline}
At lowest order in $\hbar$ (i.e., at TF level), the corresponding
single-particle propagator \cite{RingSchuck} can be written as
\begin{multline}
C^\beta(\vec{r},\vec{r}^\prime)
  = \langle\vec{r}|e^{-\beta \hat{H}}|\vec{r}^\prime\rangle
\\
  \approx \sum_\kappa \int\! \frac{d\kz\, d\kperp}{4 \pi^2}\,
    \kperp\, J_\kappa(\kperp\rperp)\,J_\kappa(\kperp\rperpprime)\,
    e^{i\kappa(\varphi-\varphiprime)}
\\
  \times e^{i\kz(z-\zprime)}\,
    e^{-\beta[V(\vec{R})+\hbar^2(\kperp^2+\kz^2)/(2m)]}\,,
\label{cbeta}
\end{multline}
where $\vec{R} = (\vec{r}+\vec{r}^\prime)/2$. Eq. (\ref{cbeta}) has
been obtained under the assumption that all the gradients of the
potential can be neglected, which is the usual hypothesis of the TF
theory. From now on we restrict ourselves to some given value of
$\kappa$. The spectral density matrix is easily obtained as the
inverse Laplace transform of the propagator \cite{RingSchuck}:
\begin{multline}
g_\kappa^\mu(\vec{r},\vec{r}^\prime) = \mathcal{L}_{\beta\to\mu}^{-1}
  C_\kappa^\beta(\vec{r},\vec{r}^\prime)
\\
  = \int\!\frac{d\kz\, d\kperp}{4\pi^2}\, \kperp\,
    J_\kappa(\kperp\rperp)\, J_\kappa(\kperp\rperpprime)\,
    e^{i\kappa(\varphi-\varphiprime)}
\\
  \times e^{i\kz(z-\zprime)}\, 
    \delta\Big(\mu-V(\vec{R})-\frac{\hbar^2(\kperp^2+\kz^2)}{2m}\Big)\, .
\label{gkappamu}
\end{multline}
Its local part, $g_\kappa^\mu(\vec{r})\equiv
g_\kappa^\mu(\vec{r},\vec{r})$, is proportional to the density of the
Bose condensate. After performing the $\kperp$ integral we obtain for
the density
\begin{multline}
\rho_\kappa(\vec{r}) = N c_\kappa g_\kappa^\mu(\vec{r})\\
  = \frac{m N c_\kappa}{2\pi^2\hbar^2}\int_0^{k_0(\vec{r})}\!\!\! dk_z\,
    J_\kappa^2\Big(\sqrt{k_0^2(\vec{r})-k_z^2}\,\rperp\Big)\,
    \theta[\mu-V(\vec{r})]\,,
\label{rhokappa}
\end{multline}
where
\begin{equation}
k_0(\vec{r}) = \frac{\sqrt{2m[\mu-V(\vec{r})]}}{\hbar}\,,
\end{equation}
and $c_\kappa$ is the normalization constant.

As it is done for the ground state \cite{SchuckVinas}, $c_\kappa$ is
determined by imposing that (\ref{rhokappa}) be normalized to
$N$. Thus $c_\kappa$ is just the inverse of the level density
$g_\kappa(\mu)$:
\begin{equation}
\frac{1}{c_\kappa} = g_\kappa(\mu) = \int\! d^3 r\, g_\kappa^\mu(\vec{r}).
\label{normalization}
\end{equation}
The other quantity entering in the density of the Bose condensate,
Eq. (\ref{rhokappa}), is the chemical potential $\mu$ which
corresponds to the lowest eigenvalue of the GPE. In order to determine
the chemical potential $\mu$, a requantization of the TF approximation
is necessary \cite{SchuckVinas}. The need for a requantization of the
TF theory for individual states has been recognized in Ref.
\cite{Krivine} and our procedure of requantization clearly follows
what is proposed there. The standard semiclassical quantization
procedure is given by the Wentzel-Kramers-Brillouin (WKB) method.
However, in order to have a more explicit formula, we apply here the
simplified method described in Ref. \cite{SchuckVinas}, which
becomes exact in the three dimensional HO case. Thus for the
non-interacting case we fix the chemical potential to be equal to the
GPE eigenvalue, Eq. (\ref{muho}).

To proceed further it is useful to write the Bessel function in Eq.
(\ref{rhokappa}) as a power series \cite{Abramowitz}:
\begin{equation}
J_\kappa(x) = \sum_{i=0}^\infty \frac{(-1)^i}{i!(\kappa+i)!}
  \Big(\frac{x}{2}\Big)^{\kappa+2i}\,.
\label{besselseries}
\end{equation}
Using this result, performing the remaining $k_z$ integral, and
remembering the identity
\begin{equation}
\sum_{i_1,i_2=0}^\infty \Big(\begin{array}{c}\kappa+j\\i_1\end{array}\Big)
        \Big(\begin{array}{c}\kappa+j\\i_2\end{array}\Big)\delta_{j,i_1+i_2}
  = \Big(\begin{array}{c}2\kappa+2j\\j\end{array}\Big)\,,
\end{equation}
we obtain the following expression for the local spectral density:
\begin{equation}
g_\kappa^\mu(\vec{r}) = \frac{m k_0(\vec{r})}{2\pi^2\hbar^2}
  \sum_{j=0}^\infty
  \frac{(-1)^j\, [k_0(\vec{r})\rperp]^{2\kappa+2j}}
    {j!(2\kappa+j)!(2\kappa+2j+1)}\theta[\mu-V(\vec{r})]\,.
\label{gkappamuseries}
\end{equation}
For the non-interacting harmonic oscillator the integral in Eq.
(\ref{normalization}) can be evaluated analytically, with the result
\begin{equation}
\frac{1}{c_\kappa} = \frac{1}{\hbar \omega}
  \sum_{j=0}^\infty \frac{(-1)^j\, [\mu/(\hbar\omega)]^{2\kappa+2j+2}}
    {j!(2\kappa+j)!(2\kappa+2j+1)(2\kappa+2j+2)}\,.
\end{equation}

\begin{figure}
\begin{center}
\epsfig{file=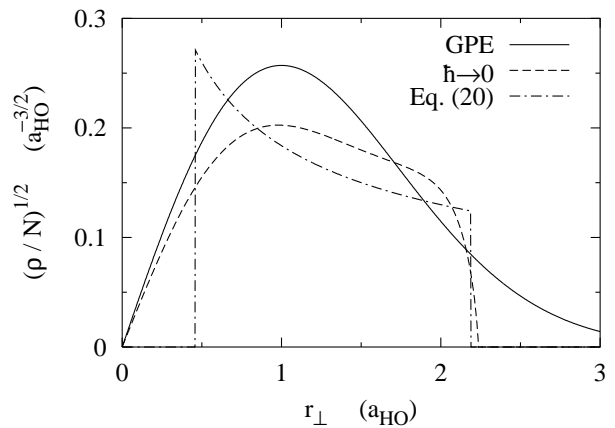,width=8.2cm}
\end{center}
\caption{\label{FigphiHO} Square root of the normalized TF ($\hbar\to
0$) density and quantal wave function (GPE) for a non-interacting Bose
condensate with $\kappa = 1$ as a function of $\rperp$ for $z =
0$. The third curve corresponds to Eq. (\ref{rhohasse}). Wave function
and radius are given in HO units ($\aHO^{-3/2}$ and $\aHO$,
respectively).}
\end{figure}

Fig. \ref{FigphiHO}
%
%%%%%%%%%%%%%%%%%%%%%%%%%%%%%%%%%%%%%%%%%%%%%%%%%%%%%%%%%%%%%%%%%%%%%%%%%
% Figure 1
%
%%%%%%%%%%%%%%%%%%%%%%%%%%%%%%%%%%%%%%%%%%%%%%%%%%%%%%%%%%%%%%%%%%%%%%%%%
%
displays the square root of the normalized TF density (\ref{rhokappa})
for $\kappa=1$ along the $\rperp$ coordinate for $z = 0$, where HO
units have been used. In the same figure the quantal wave function
which describes the $\kappa = 1$ vortex state [see Eq.
(\ref{solutionho})] is also plotted. As it can be seen from Eq.
(\ref{gkappamuseries}), for $\rperp\to 0$ the semiclassical TF density
goes to zero in the same way as the quantum mechanical result, i.e.,
$\rho_\kappa(\vec{r})\propto \rperp^{2\kappa}$. At the classical
turning point [$V(\vec{r}) = \mu$] the TF density goes to zero as
$\rho_\kappa\propto[k_0(\vec{r})]^{1+2\kappa}$.  Thus the turning point
will be changed by the interaction only via the change of the chemical
potential $\mu$.

The reader unfamiliar with the Thomas-Fermi approach may be worried
about the locally relatively strong deviations of the semiclassical
density from its quantal (GPE) counterpart. In this respect it should
be remembered that the TF densities must be considered in the sense of
distributions \cite{RingSchuck} [see e.g. the step function in Eq.
(\ref{rhokappa})] and, therefore, they only make sense when used under
integrals to calculate expectation values of ``slowly varying''
operators. In fact, the $\hbar$ corrections to the density (not
considered here) still deviate much more strongly from the true
quantal densities, since they contain a square-root singularity at the
classical turning point. Nonetheless these $\hbar$ corrections, when
used under integrals, improve the results for cases where the
gradients of the potential are not too strong \cite{Centelles}. It has
been found in the past \cite{SchuckVinas,RingSchuck} that when used in
this way the TF approach (eventually with inclusion of $\hbar$
corrections) can yield very accurate results for expectation
values. The relation of the TF approach with $\hbar^2$ corrections to
the WKB approach has been discussed in Ref. \cite{Krivine}.

The TF kinetic energy density can also be derived from the spectral
density matrix (\ref{gkappamu}) as
\begin{align}
\tau_\kappa(\vec{R})
  = &\Big(-\frac{\hbar^2}{2m} \vec{\nabla}_r^2
    g_\kappa^\mu(\vec{r},\vec{r}^\prime)\Big)_{\vec{r},\vec{r}^\prime
      \to\vec{R}} \nonumber\\
  = &\frac{\hbar^2}{2m}\int\!\frac{dk_z\, d\kperp}{4\pi^2} \kperp
    (\kperp^2+k_z^2)\, J_\kappa^2(\kperp\Rperp) \nonumber\\
    &\qquad\qquad\times\delta\Big(\mu-V(\vec{R})
       -\frac{\hbar^2}{2m}(\kperp^2+k_z^2)\Big)
      \nonumber\\
  = &[\mu-V(\vec{R})]\, g_\kappa^\mu(\vec{R}).
\label{tau}
\end{align}
Using Eqs. (\ref{rhokappa}) and (\ref{tau}) it can easily be checked
that the expectation values of the kinetic and potential energies
fulfill the virial theorem as it is expected in the TF approach
\cite{Englert}. Thus due to our choice of the chemical potential $\mu$
(\ref{muho}), our semiclassical TF approximation to the vortex state
in the non-interacting case exactly reproduces the quantal expectation
values of the kinetic and potential energies in spite of the aspect of
the semiclassical density profile as compared with the quantal one.

It is easy to see that with the HO potential the argument of the
Bessel function entering in the density cannot become large:
\begin{equation}
\kperp\rperp\leq k_0(\vec{r})|\vec{r}|\leq\frac{\mu}{\hbar\omega}
  = \frac{3}{2}+\kappa\,.
\end{equation}
For example, in the case $\kappa = 1$ the argument becomes at most
$5/2$, and already the first four terms of the expansion
(\ref{besselseries}) give an accuracy better than $0.5\%$. In the case
$\kappa = 0$, the result of Ref. \cite{SchuckVinas} is recovered if
one takes only the first term of the expansion in the TF density,
Eq. (\ref{rhokappa}).

For completeness we note that in the literature also a different
approach for projecting the semiclassical density matrix onto good
angular momentum $L^2$ and $L_z$ can be found \cite{Hasse}. Repeating
the steps described there for the projection onto good $L_z$ only
(i.e., essentially using asymptotic expansions for the Bessel
functions) one finds the following expression for the Wigner transform
of the spectral density matrix:
\begin{equation}
g_\kappa^\mu(\vec{R},\vec{p}) = \hbar\,
  \delta(H^\mathit{cl}-\mu)\, \delta(L_z^\mathit{cl}-\hbar \kappa)
\end{equation}
with $H^\mathit{cl} = \vec{p}^2/(2m)+V(\vec{R})$ and
$\vec{L}^\mathit{cl} = \vec{R}\times\vec{p}$\,. From this formula the density
is easily obtained by integration over $\vec{p}$:
\begin{equation}
\rho_\kappa(\vec{R})=
  \frac{m Nc_\kappa}{4\pi^2\hbar^2 \Rperp}\,
  \theta\Big(\mu-V(\vec{R})-\frac{\hbar^2\kappa^2}{2m\Rperp^2}\Big)\,.
\label{rhohasse}
\end{equation}
The constant $c_\kappa$ is determined by the normalization condition,
which for the non-interacting case results in $c_\kappa =
2\hbar^2\omega^2/(\mu-\kappa\hbar\omega) = 4\hbar\omega/3$\,.  The
density profile corresponding to Eq. (\ref{rhohasse}) is also shown
in Fig. \ref{FigphiHO}. From this figure it is evident that Eq.
(\ref{rhohasse}) makes sense only as a distribution for the
calculation of expectation values and not for the calculation of local
quantities like the density itself. However, since the density given
by Eq. (\ref{rhohasse}) does not at all depend on the shape of the
potential (except for the determination of the turning points), this
form seems difficult to be used for a self-consistent calculation in
the interacting case. Let us note that again the virial theorem is
fulfilled and expectation values of operators can be obtained very
accurately \cite{Hasse,VinasSchuck}.
%
%%%%%%%%%%%%%%%%%%%%%%%%%%%%%%%%%%%%%%%%%%%%%%%%%%%%%%%%%%%%%%%%%%%%%%%%%
\section{The interacting case}
Let us now discuss the TF approximation to the quantal solution of the
GPE (\ref{gpe}). Of course the semiclassical formalism described in
the previous section can still be applied provided that the potential
$V(\vec{r})$ in the interacting case is given by
\begin{equation}
V(\vec{r}) = \Vext(\vec{r}) + g\rho(\vec{r})\,.
\end{equation}
As it was mentioned before, the TF density corresponding to the vortex
state (\ref{rhokappa}) depends on two independent constants to be
determined: the normalization $c_\kappa$ and the chemical potential
$\mu$. In the interacting case, and following the same strategy as in
Ref. \cite{SchuckVinas}, we determine $c_\kappa$ and $\mu$ by
imposing that the TF density be normalized to the number of particles
$N$ in the Bose condensate and that the integrated level density
\begin{align}
N_\kappa^I(\mu) =& \int\! d^3 r\int_0^\mu \!\! d\muprime\,
  g_\kappa^\muprime(\vec{r})\nonumber\\
=& \int \! d^3 r\, \theta[\mu-V(\vec{r})]\,\frac{k_0^3(\vec{r})}{2\pi^2}
  \nonumber\\
 &\times \sum_{j=0}^\infty
  \frac{(-1)^j\, [k_0(\vec{r}) \rperp]^{2\kappa+2j}}
  {j!(2\kappa+j)!(2\kappa+2j+1)(2\kappa+2j+3)} 
\label{quantcond}
\end{align}
become equal to that of the non-interacting HO, which for $\mu =
(3/2+\kappa)\hbar \omega$ is given by
\begin{equation}
N_\kappa^{\mathit{HO}} = \sum_{j=1}^{\infty}
  \frac{(-1)^j\, (2\kappa+2j)!\,(3/2+\kappa)^{2\kappa+2j+3}}
    {j!(2\kappa+j)!(2\kappa+2j+3)!}\,.
\label{quantcondho}
\end{equation}

The strategy for the self-consistent solution in the interacting case
is now very simple. Instead of starting with a fixed particle number
$N$, it is convenient to choose some value for $N c_\kappa$. Then we
choose some initial value for $\mu$. For given $Nc_\kappa$ and $\mu$
we solve Eq. (\ref{gkappamuseries}) for $g_\kappa^\mu(\vec{r})$, which
is non-linear since also the right-hand side depends on
$g_\kappa^\mu(\vec{r})$ through
\begin{equation}
k_0(\vec{r}) = \frac{\sqrt{2m[\mu-\Vext(\vec{r})-gNc_\kappa
  g_\kappa^\mu(\vec{r})]}}{\hbar}\,.
\end{equation}
Then the integral (\ref{quantcond}) is evaluated and the result is
compared with the corresponding result for the non-interacting
harmonic oscillator, Eq. (\ref{quantcondho}). If the level number is
too small ($N_\kappa^I < N_\kappa^{\mathit{HO}}$), $\mu$ is increased,
otherwise ($N_\kappa^I > N_\kappa^{\mathit{HO}}$) $\mu$ is
decreased. This procedure is iterated until $N_\kappa^I =
N_\kappa^{\mathit{HO}}$. Finally the particle number is obtained by
evaluating the integral
\begin{equation}
N = \int\! d^3r\, \rho(\vec{r}) = N c_\kappa \int\! d^3r\,
  g_\kappa^\mu(\vec{r})\, .
\end{equation}

Before comparing the results obtained within our TF approach to the
results from solving the GPE numerically, let us briefly discuss two
approximation methods which have been developed for the case of large
$N$. The first one, known as the TF limit in the literature and
discussed, e.g., in Refs. \cite{Fetter,Rokhsar,Ho,Sinha}, is obtained
by dropping the kinetic energy part $e_\mathit{kin}$ coming from the
radial and axial motion and retaining only the rotational part
$e_\mathit{rot}$ of the total kinetic energy, i.e., only derivatives
with respect to $\varphi$ in Eq. (\ref{gpe}). Under this assumption it
is easily obtained that the $N\to\infty$ limit of the density of the
vortex state reads
\begin{equation}
\rho(\vec{r}) = \frac{1}{g}\Big(\mu-\frac{\hbar^2\kappa^2}{2m\rperp^2}
  -\frac{1}{2} m\omega^2 (\rperp^2+z^2)\Big)\, .
\label{rhorokhsar}
\end{equation}
In this limit the density vanishes inside of $\rperpmin$ and outside
of $\rperpmax$, defined by the zeros of Eq. (\ref{rhorokhsar}). The
chemical potential $\mu$ is obtained through the particle-number
condition. The formula (\ref{rhorokhsar}) has the advantage that it
represents an analytic expression for $\rho(\vec{r})$, but it is clear
that the result $\rho = 0$ inside a vortex core with radius
$\rperpmin$ is not realistic. We will call hitherto formula
(\ref{rhorokhsar}) the $N\to\infty$ TF limit.

The second approximation method, known as the ``method of matched
asymptotics'' (MA), was introduced in Ref. \cite{Svidzinsky} to
describe the dynamics of vortices, and used in Ref. \cite{Aftalion} to
calculate the energy of a static vortex. We will follow here the
simplified derivation for the case of a straight vortex given in
Ref. \cite{Modugno}. First let us briefly review the description of a
vortex state in a system with $\Vext = 0$. In this case it is useful
to define the asymptotic density $\rho_0 = \mu/g$ and the healing
length $\xi_0 = \hbar/\sqrt{2m\rho_0 g}$ \cite{Fetter}, and to write
the condensate wave function in the form
\begin{equation}
\phi(\rperp) = \sqrt{\rho_0}\, f_\kappa\Big(\frac{\rperp}{\xi_0}\Big)
  = \sqrt{\frac{\mu}{g}}\,
    f_\kappa\Big(\frac{\sqrt{2m\mu}}{\hbar}\,\rperp\Big)\, .
\label{phihomogen}
\end{equation}
Inserting this expression into the GPE (\ref{gpe}) with $\Vext = 0$,
one obtains the following differential equation for the function
$f_\kappa$:
\begin{equation}
-\frac{1}{x} f_\kappa^\prime(x)-f_\kappa^{\prime\prime}(x)+
  \frac{\kappa^2}{x^2}f_\kappa(x)+f_\kappa^3(x)=f_\kappa(x)\,.
\label{fdgl}
\end{equation}
With the boundary conditions (for $\kappa \geq 1$)
\begin{equation}
f_\kappa(0) = 0 \qquad \mbox{and}
  \qquad \lim_{x\to\infty}f_\kappa(x) = 1
\label{fboundary}
\end{equation}
this differential equation can be solved numerically
\cite{Fetter,Ginzburg}. Now we turn to the case of a trapped system.
In the limit of large $N$ it is is clear that the external potential
$\Vext$ can be regarded as constant on the length scale $\xi_0$
corresponding to the size of the vortex core. (More precisely, the
condition which has to be fulfilled reads $Na/\aHO \gg 1$, as it is
the case for the $N\to\infty$ TF approach.) Thus we obtain an
approximate description of the trapped system by replacing the
chemical potential $\mu$ by a local chemical potential
$\mu-\Vext(\vec{r})$. Inside the classically allowed region
[$\Vext(\vec{r})<\mu$] the order parameter then takes the form
\begin{equation}
\phi(\vec{r}) = \sqrt{\frac{\mu-\Vext(\vec{r})}{g}}\,
  f_\kappa\Big(\frac{\sqrt{2m[\mu-\Vext(\vec{r})]}}{\hbar}\,r_\perp\Big)\,.
\label{philocal}
\end{equation}
As before, the chemical potential $\mu$ is determined by the
particle-number condition.

Note that in the region far away from the vortex core, i.e., for
$\rperp\gg\xi_0$, one can expand $f_\kappa(x)$ in powers of
$1/x$. Using Eqs. (\ref{fdgl}) and (\ref{fboundary}), one obtains
\begin{equation}
f_\kappa(x) = 1-\frac{\kappa^2}{2x^2}-\frac{\kappa^2(8+\kappa^2)}{8x^4}-\dots
  \approx \sqrt{1-\frac{\kappa^2}{x^2}}\,.
\label{fasymptot}
\end{equation}
Inserting this into Eq. (\ref{philocal}) one immediately recovers
Eq. (\ref{rhorokhsar}). However, this shows that Eq.
(\ref{rhorokhsar}) is not valid for $\rperp\lesssim \xi_0$, i.e.,
inside the vortex core. Another difference between
Eq. (\ref{philocal}) and Eq. (\ref{rhorokhsar}) concerns the behavior
of the wave function at the outer classical turning point. In contrast
to the usual $N\to\infty$ TF limit, the kinetic energy corresponding
to the wave function (\ref{philocal}) is not diverging, since the
square-root $\sqrt{\mu-\Vext}$ in Eq. (\ref{philocal}) is
multiplied by the function $f_\kappa$, which is proportional to
$(\mu-\Vext)^{\kappa/2}$ near the classical turning
point. Nevertheless it is not reasonable to use Eq. (\ref{philocal})
to calculate the kinetic energy near the turning point, since the
decrease of the function $f_\kappa$ is just indicating that the local
healing length $\xi(\vec{r}) = \hbar/\sqrt{2m(\mu-\Vext)}$
becomes large and that the approximation breaks down.

Now we proceed to a detailed numerical comparison of the ($\hbar\to
0$) TF predictions with the exact quantal values obtained from the
GPE, Eq. (\ref{gpe}). For our numerical application we consider
$^{87}$Rb atoms in a spherical trap represented by a HO potential with
length $\aHO = 0.791$ $\mu$m \cite{Dalfovo99}. The $s$-wave scattering
length is taken as $a = 100\, a_0$ \cite{Dalfovo96a} where $a_0$ is
the Bohr radius. Table \ref{table1}
%
%%%%%%%%%%%%%%%%%%%%%%%%%%%%%%%%%%%%%%%%%%%%%%%%%%%%%%%%%%%%%%%%%%%%
% Table I
%
\begin{table}
\newcommand{\notdef}{\multicolumn{1}{c}{---}}
\newcommand{\neglect}{\multicolumn{1}{c}{0}}
\newcommand{\rhorokhsar}{Eq. (\ref{rhorokhsar})}
\newcommand{\philocal}{Eq. (\ref{philocal})}
\begin{tabular}{llrrrrrr}
\hline\noalign{\smallskip}
$N$ & &\multicolumn{1}{c}{$\mu$} & \multicolumn{1}{c}{$\etot$} &
\multicolumn{1}{c}{$\eHO$} & \multicolumn{1}{c}{$\eself$} &
\multicolumn{1}{c}{$\erot$} & \multicolumn{1}{c}{$\ekin$} \\
\noalign{\smallskip}\hline\noalign{\smallskip}
 & GPE          &  2.74 &    2.62 &  1.34 &  0.12 & 0.480 &   0.686  \\
 & $\hbar\to 0$ &  2.72 &    2.61 &  1.33 &  0.10 & 0.501 &   0.673  \\
\raisebox{1.5ex}[0pt]{$10^2$}
 & $N\to\infty$ &  1.88 &    1.59 &  0.87 &  0.29 & 0.438 & \neglect \\
 & MA           &  1.86 & \notdef &  0.84 &  0.23 & 0.689 & \notdef  \\
\noalign{\smallskip}\hline\noalign{\smallskip}
 & GPE          &  8.40 &   6.30  &  3.67 &  2.10 & 0.271 &   0.255  \\
 & $\hbar\to 0$ &  8.28 &   6.19  &  3.62 &  2.09 & 0.350 &   0.130  \\
\raisebox{1.5ex}[0pt]{$10^4$}
 & $N\to\infty$ &  8.19 &   5.99  &  3.54 &  2.19 & 0.253 & \neglect \\
 & MA           &  8.23 & \notdef &  3.57 &  2.17 & 0.272 & \notdef  \\
\noalign{\smallskip}\hline\noalign{\smallskip}
 & GPE          & 50.18 &   35.93 & 21.53 & 14.26 & 0.087 &   0.059  \\
 & $\hbar\to 0$ & 50.13 &   35.86 & 21.50 & 14.27 & 0.116 &  -0.024  \\
\raisebox{1.5ex}[0pt]{$10^6$}
 & $N\to\infty$ & 50.14 &   35.86 & 21.50 & 14.28 & 0.083 & \neglect \\
 & MA           & 50.14 & \notdef & 21.50 & 14.27 & 0.086 & \notdef  \\
\noalign{\smallskip}\hline
\end{tabular}
\caption{\label{table1} Chemical potential ($\mu$) and energy per
particle ($\etot$) and its different contributions in $\hbar \omega$
units: harmonic oscillator energy ($\eHO$), interaction energy
($\eself$), and kinetic energy split into its rotational ($\erot$) and
radial and axial ($\ekin$) parts. The parameters chosen correspond to
a single-quantized vortex ($\kappa = 1$) in a spherical trap ($\aHO =
0.791$ $\mu$m) containing $100$, $10^4$, and $10^6$ $^{87}$Rb atoms
(scattering length $a = 100\,a_0$). The results obtained from the GPE
are compared with the results from the ($\hbar\to 0$) TF approach and
from two approximation methods for large $N$: the so-called
$N\to\infty$ TF method, Eq. (\ref{rhorokhsar}), and the method of
matched asymptotics (MA), Eq. (\ref{philocal}). Note that $\ekin$ is
neglected in the $N\to\infty$ TF limit and not accessible within the
matched-asymptotic approach.}
\end{table}
%%%%%%%%%%%%%%%%%%%%%%%%%%%%%%%%%%%%%%%%%%%%%%%%%%%%%%%%%%%%%%%%%%%%%%%%
%
collects the chemical potential ($\mu$), the total ($\etot$), HO
($\eHO$), self-interaction ($\eself$), and kinetic energies per
particle for vortex states of condensates with $100$, $10^4$, and
$10^6$ atoms in the trap. The kinetic energy is split into the
rotational part ($\erot$) and in the one corresponding to the radial
and axial motion ($\ekin$). The numerical values displayed in Table
\ref{table1} show that our ($\hbar\to 0$) TF approach reproduces very
well the quantal eigenvalue ($\mu$) as well as the total energy per
particle ($\etot$) even for a small number of particles such as
100. The agreement between the quantal and TF values improves when the
number of particles in the condensate increases, as it is
expected. The HO and the self-interaction contributions to the total
energy are also well reproduced by our semiclassical approach. For
very large numbers of particles ($N = 10^6$) the quantal results are
also well reproduced by the $N\to\infty$ TF limit,
Eq. (\ref{rhorokhsar}), because the neglected contribution (i.e., the
kinetic energy due to the radial and axial motion $\ekin$) is very
small. However, it should be pointed out that the key assumption of
this $N\to\infty$ limit is not fulfilled, because the kinetic energy
of the radial and axial motion is still of the same order as the
rotational energy, as can be seen from the quantal results (GPE)
listed in Table \ref{table1}. In fact, even in the limit $N\to\infty$
the ratio $\ekin/\erot$ does not go to zero (see appendix). The method
of matched asymptotics, Eq. (\ref{philocal}), gives better results
than the $N\to\infty$ TF limit, Eq.  (\ref{rhorokhsar}), except in the
case of small numbers of particles ($N = 100$), where both large-$N$
methods fail.

Concerning the kinetic energy some comments are in order. First of
all, we want to point out that the $N\to\infty$ theory neglects (and
in fact cannot access \cite{SchuckVinas}) the contributions coming
from the radial and axial motion, so they are not listed in Table
\ref{table1}. For a small number of atoms, such as 100, our $\hbar\to
0$ limit is able to reproduce reasonably well both, $\erot$ and
$\ekin$ contributions to the total kinetic energy per particle. When
the number of the atoms in the trap grows, the total kinetic energy
per particle decreases and the agreement between the quantal result
and the TF prediction worsens for this quantity. This situation is
also found in the ground-state case discussed in Ref.
\cite{SchuckVinas} where the (small) quantal and TF kinetic energies
can differ by a factor two for a large number of particles (see Table
II of Ref. \cite{SchuckVinas}). The reason for these disagreements
between the quantal and TF kinetic energies for large number of atoms
in the condensate lies in the fact that in this case the kinetic
energy is dominated by quantal corrections that are non-analytical in
$\hbar$ and consequently cannot be reproduced in a pure TF
approximation \cite{SchuckVinas}. A detailed comparison shows that the
($\hbar\to 0$) TF theory systematically overestimates the rotational
part and underestimates the axial and radial parts of the kinetic
energy, the latter even becoming negative for very large numbers of
particles, although the total kinetic energy remains always
positive. The reason for this behavior is that the TF density is too
high inside the vortex core, as will be discussed below.

It should be pointed out that, as happens for the non-interacting case,
the virial theorem, which for the interacting case reads
\cite{Dalfovo96a}
\begin{equation}
2(\ekin+\erot)-2\eHO+3\eself = 0\,,
\end{equation}
is also fulfilled in our TF approach to vortex states for a Bose
condensate in a spherical trap.

\begin{figure}
\begin{center}
\epsfig{file=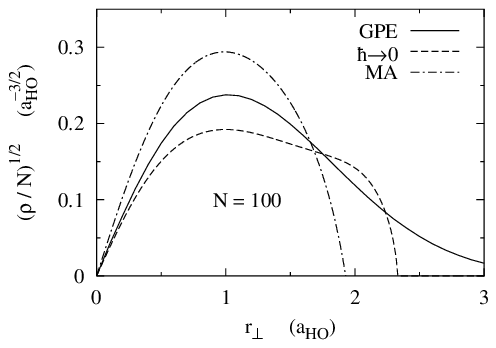,width=8.2cm}
\end{center}
\caption{\label{FigphiRb2} Normalized order parameter obtained from
the GPE, from the ($\hbar\to 0$) TF limit, and from the approximation of
matched asymptotics (MA) for large $N$ [Eq. (\ref{philocal})],
of an interacting Bose condensate of $100$ $^{87}$Rb atoms in a
spherical trap with $\aHO = 0.791$ $\mu$m in a vortex state with
$\kappa = 1$ as a function of $\rperp$ for $z = 0$ in HO units.}
\begin{center}
\epsfig{file=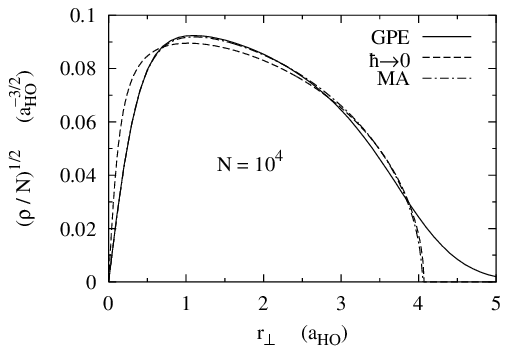,width=8.2cm}
\end{center}
\caption{\label{FigphiRb4} Same as Fig. \ref{FigphiRb2}, but for
$10^4$ atoms in the trap.}
\begin{center}
\epsfig{file=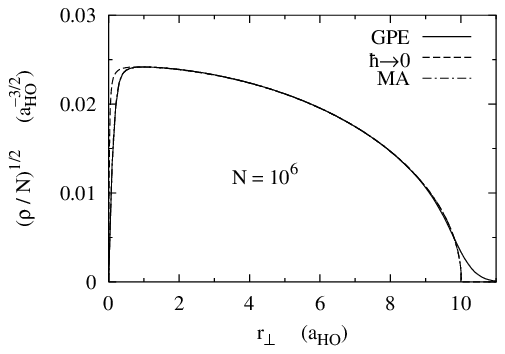,width=8.2cm}
\end{center}
\caption{\label{FigphiRb6} Same as Fig. \ref{FigphiRb2}, but for
$10^6$ atoms in the trap.}
\end{figure}

Figures \ref{FigphiRb2}-\ref{FigphiRb6}
%
%%%%%%%%%%%%%%%%%%%%%%%%%%%%%%%%%%%%%%%%%%%%%%%%%%%%%%%%%%%%%%%%%%%%%%%%%
% Figures 2-4
%
%%%%%%%%%%%%%%%%%%%%%%%%%%%%%%%%%%%%%%%%%%%%%%%%%%%%%%%%%%%%%%%%%%%%%%%%%
%
display the normalized order parameter for $100$, $10^4$, and $10^6$
atoms of $^{87}$Rb in the trap along the radial axis $\rperp$ for
$z=0$. The dashed line corresponds to the ($\hbar\to 0$) TF limit.
For comparison we show the corresponding order parameter obtained from
the quantal solution of the GPE (\ref{gpe}) (solid line), which is
obtained through imaginary time step techniques \cite{Dalfovo96a}, and
the order parameter obtained from the method of matched asymptotics,
Eq. (\ref{philocal}) (dashed-dotted lines). Looking at the shape of
the semiclassical ($\hbar\to 0$) compared with the quantal order
parameter one can see that the agreement increases with the number of
particles in the condensate, as it happens in the TF approximation for
the ground state. The effect of the self-interaction that
progressively modifies the density profile of the condensate in the
vortex state with respect to the non-interacting case is also followed
by our semiclassical TF densities. Only inside the vortex core
($\rperp\approx 0$) the agreement worsens with increasing number of
particles.

\begin{figure}
\begin{center}
\epsfig{file=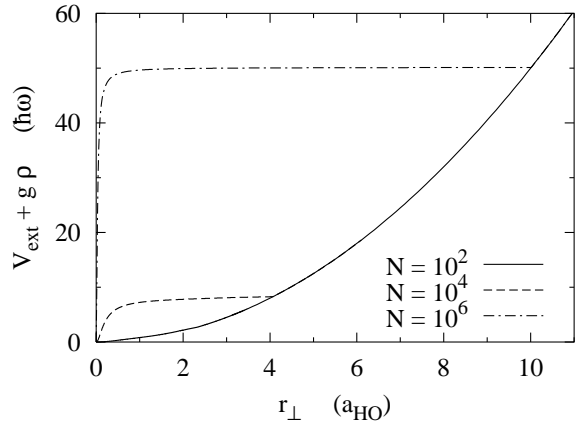,width=8.2cm}
\end{center}
\caption{\label{FigVsc} Self-consistent potentials in units of
$\hbar\omega$ as a function of $\rperp$ in units of $\aHO$ obtained in
our ($\hbar\to 0$) TF approach corresponding to the density profiles
shown in Figs. \ref{FigphiRb2}-\ref{FigphiRb6}.}
\end{figure}

This can easily be understood by looking at the corresponding
self-consistent potentials shown in Fig. \ref{FigVsc}.
%
%%%%%%%%%%%%%%%%%%%%%%%%%%%%%%%%%%%%%%%%%%%%%%%%%%%%%%%%%%%%%%%%%%%%%%%%%
% Figure 5
%
%%%%%%%%%%%%%%%%%%%%%%%%%%%%%%%%%%%%%%%%%%%%%%%%%%%%%%%%%%%%%%%%%%%%%%%%%
%
The main assumption of our semiclassical TF theory is that gradients
of the potential can be neglected. This assumption becomes more and
more justified with increasing number of particles, except in the
vicinity of the $z$ axis ($\rperp\approx 0$), where the self
consistent potential rises rapidly from zero to $\approx\mu$. For the
case of moderate numbers of particles, the semiclassical description
of the vortex core could be improved by considering higher $\hbar$
corrections to the TF solution, which take into account the gradients
of the potential. However, we should remember that the $\hbar$ or
gradient expansion is an asymptotic series which can only work as long
as the gradients of the potential are not too strong, even though the
theory often works quite far beyond its limits (see Ref.
\cite{Centelles}). For very steep potentials only a partial
resummation of the $\hbar$ series like in WKB, to account for the
nonanalytical behavior in $\hbar$, can help. This will further be
discussed in the Appendix.

As can be seen in Figs. \ref{FigphiRb4} and \ref{FigphiRb6}, for large
numbers of particles the density profiles obtained from Eq.
(\ref{philocal}) follow remarkably well the quantal profile except
near the classical turning point where the approach breaks down.

Comparing the so-called $N\to\infty$ TF approach [Eq.
\ref{rhorokhsar})] with the $\hbar\to 0$ TF approach proposed in this
work, one can see from Table \ref{table1} that our TF method
reproduces better the different quantal contributions to the energy of
the vortex state for small ($N = 100$) and moderate ($N = 10^4$)
numbers of particles in the condensate while, for large numbers both
limits ($\hbar\to 0$ and $N\to\infty$) coincide. Concerning the
density profiles, the difference is obvious: In our approach the
density profile goes like $\sqrt{\rho}\propto\rperp^\kappa$ as in the
quantal case, whereas in the $N\to\infty$ TF limit there is a small
$\rho = 0$ region determined by the inner turning point $\rperpmin$.
It should be mentioned that within the improved (matched asymptotics)
$N\to\infty$ approximation [Eq. (\ref{philocal})], the density profile
also goes like $\sqrt{\rho}\propto\rperp^\kappa$, and contrary to our
$\hbar\to 0$ approach it is capable to reproduce the density profile
in the vortex core in the case of large $N$. However, this method has
nothing to do with the semiclassical asymptotic $\hbar$ expansion
considered here.

Finally it should also be pointed out that our TF limit is able to
deal with vortices in the attractive case (negative scattering
length). In this case the kinetic energy is crucial and the large-$N$
limit is not well-defined. The same is true for the description of the
ground state (i.e., no vortex), as shown in
Ref. \cite{SchuckVinas}. There the $\hbar\to 0$ approach has been used
in the repulsive as well as in the attractive case, whereas the
$N\to\infty$ approximations (the so-called $N\to\infty$ TF limit as
well as the method of matched asymptotics) can be applied only in the
repulsive case.
%
%%%%%%%%%%%%%%%%%%%%%%%%%%%%%%%%%%%%%%%%%%%%%%%%%%%%%%%%%%%%%%%%%%%%%
\section{Application to vortices in superfluid trapped fermionic
  gases}
In this section we will describe how our TF approach can also be used
for the description of vortices in superfluid fermionic systems. This
is possible since at least for a certain range of temperatures, the
so-called Ginzburg-Landau regime, the order parameter $\Delta(\vec{r})$
is described by an equation which has exactly the same form as the GPE
(\ref{gpe}). As derived in Ref. \cite{Baranov}, for temperatures $T$
near the critical temperature $T_c$ and for low trapping frequencies
$\omega$ ($\hbar\omega\ll k_B T_c$) the Ginzburg-Landau equation (GLE)
reads
\begin{multline}
\Big[-K^2 R_{\mathit{TF}}^2 \vec{\nabla}^2
     +\frac{1+2\lambda}{2\lambda}\frac{\vec{r}^2}{R_{\mathit{TF}}^2}
     -\ln\frac{T_c^{(0)}}{T}\Big]\Delta(\vec{r})\\
 +\frac{7\zeta(3)}{8\pi^2}\left|\frac{\Delta(\vec{r})}{k_B T}\right|^2
   \Delta(\vec{r}) = 0\, ,
\label{gle}
\end{multline}
with the definitions $K = \sqrt{7\zeta(3)/(48\pi^2)}\,\hbar\omega/(k_B
T)$, $\lambda = 2 k_F|a|/\pi$, and $R_{\mathit{TF}} = \hbar
k_F/(m\omega)$, where $k_F$ denotes the local Fermi momentum at the
center of the trap. The temperature $T_c^{(0)}$ is the critical
temperature of a homogeneous system having the same density as the
trapped system has at the center.  It is given by $T_c^{(0)} = (8
e^{-2}\gamma/\pi) \epsilon_F e^{-1/\lambda}$ \cite{Melo}, with $\gamma
\approx 1.781$ and $\epsilon_F = \hbar^2k_F^2/(2m)$.

It is convenient to rewrite Eq. (\ref{gle}) in terms of
dimensionless quantities. To that end we define
\begin{align}
\tilde{\vec{r}} = & \Big(\frac{1}{K}\Big)^{1/2}\,
                   \Big(1+\frac{1}{2\lambda}\Big)^{1/4}
                   \frac{\vec{r}}{R_{\mathit{TF}}}\, ,\\
\tilde{g} = & \frac{7\zeta(3)}{16\pi^2}\,\frac{1}{K}\,
              \Big(\frac{2\lambda}{1+2\lambda}\Big)^{1/2}\, ,\\
\tilde{\mu} = & \frac{1}{2K}\,\Big(\frac{2\lambda}{1+2\lambda}\Big)^{1/2}
                \ln\frac{T_c^{(0)}}{T}\, ,\label{defmutilde}\\
\tilde{\Phi} = & \frac{\Delta}{k_B T}\, .
\end{align}
With these definitions, Eq. (\ref{gle}) becomes
\begin{equation}
\Big(-\frac{1}{2}\tilde{\vec{\nabla}}^2 + \frac{1}{2}\tilde{\vec{r}}^2
  +\tilde{g}|\tilde{\Phi}(\tilde{\vec{r}})|^2\Big)
  \tilde{\Phi}(\tilde{\vec{r}})
  =\tilde{\mu}\tilde{\Phi}(\tilde{\vec{r}})\, ,
\label{gpehounits}
\end{equation}
which is the same as the GPE rewritten in HO units, i.e., with the
replacements $\vec{r}/\aHO\to \tilde{\vec{r}}$,
$g/(\hbar\omega \aHO^3)\to \tilde{g}$, and
$\phi\, \aHO^{3/2}\to\tilde{\phi}$.

However, there is one important difference between the GPE describing
the Bose-Einstein condensate and the GLE describing the order
parameter $\Delta(\vec{r})$ of a superfluid Fermi system. In a
Bose-Einstein condensate, the particle number $N$, i.e., the norm of
$\tilde{\Phi}$, is fixed, and the chemical potential $\tilde{\mu}$ has
to be determined from the GPE (\ref{gpehounits}). For the GLE the
situation is reversed: The chemical potential $\tilde{\mu}$ is fixed
by the temperature $T$ and other parameters [Eq.
(\ref{defmutilde})], whereas the normalization of $\tilde{\Phi}$,
i.e., the magnitude of the gap $\Delta$, has to be determined from
Eq. (\ref{gpehounits}).

The lowest possible value of $\tilde{\mu}$, for which a solution of
the GLE (\ref{gpehounits}) can be found, corresponds to the case that
the normalization $N$ goes to zero, such that the non-linear term can
be neglected. In this case Eq. (\ref{gpehounits}) reduces to the
Schr\"odinger equation of the non-interacting harmonic oscillator with
the lowest eigenvalue $\tilde{\mu}_{\mathit{min}} = 3/2$. This gives
an upper limit for $T/T_c^{(0)}$, which was used in Ref.
\cite{Baranov} to estimate the critical temperature $T_c$ of the
trapped Fermi system.

In this article we are interested in vortex states, i.e., in solutions
of the form (\ref{formvortex}). In the framework of the GL theory,
vortex states of superfluid Fermi systems are discussed in Ref.
\cite{Rodriguez}, where the GLE (\ref{gpehounits}) is solved for a
two-dimensional geometry, i.e., $\phi(\rperp,z)\equiv\phi(\rperp)$,
corresponding to a trap with an extremely elongated potential. In two
dimensions the lowest possible value of $\tilde{\mu}$, for which
vortex solutions can be found, is given by $\mu_{\mathit{min}}^{2d} =
1+\kappa$. However, as in our discussion of vortex states in
Bose-Einstein condensates, we will consider the spherical case, in
which the maximum temperature for the existence of vortex states is
determined by $\tilde{\mu}_{\mathit{min}} = 3/2+\kappa$.

Since the GLE is identical with the GPE, it is obvious that our TF
approach described in the previous sections can immediately be applied
also to the GLE. Only the iteration procedure for the self-consistent
solution is somewhat different, since now $\tilde{\mu}$ is given
instead of $N$. We start with some guess for $N c_\kappa$ and
calculate the integrated level density, Eq. (\ref{quantcond}). Now,
if $N_\kappa^I < N_\kappa^{\mathit{HO}}$, the value of $Nc_\kappa$ is
increased, otherwise it is decreased. This procedure is iterated until
$N_\kappa^I = N_\kappa^{\mathit{HO}}$. Due to this quantization rule
it is clear that $\Delta$ goes to zero when the temperature approaches
the critical temperature corresponding to $\tilde{\mu} = 3/2+\kappa$.

We are now going to compare the results of our TF approach with the
fully quantal solution of Eq. (\ref{gpehounits}). The parameters
used for our calculations are taken from Ref. \cite{Houbiers}, i.e.,
we consider $N_{^6\mathrm{Li}} = 573000$ $^6$Li atoms (scattering
length $a = -2160\, a_0$) in a trap with $\omega = 2\pi\times 144$
Hz. The self-consistent mean-field potential of the cloud has been
neglected in the derivation of the GLE (\ref{gle}) in Ref.
\cite{Baranov}, but we take it into account in an approximate way by
replacing the external trapping frequency $\omega$ by a higher one,
$\omega_\mathit{eff} = 2\pi\times 170$ Hz, as it has been done, e.g.,
in Ref. \cite{Farine}. The parameters $R_{\mathit{TF}}$ and
$T_c^{(0)}$ are obtained from $\epsilon_F = (3
N_{^6\mathrm{Li}})^{1/3}\hbar\omega_{\mathit{eff}}$ and the relations
given below Eq. (\ref{gle}), with the result $R_{\mathit{TF}} =
48.7$ $\mu$m and $T_c^{(0)} = 36.7$ nK. The temperature corresponding
to $\tilde{\mu} = 3/2$, i.e., the critical temperature of the trapped
system, is $T_c = 31.2$ nK.

\begin{figure}
\begin{center}
\epsfig{file=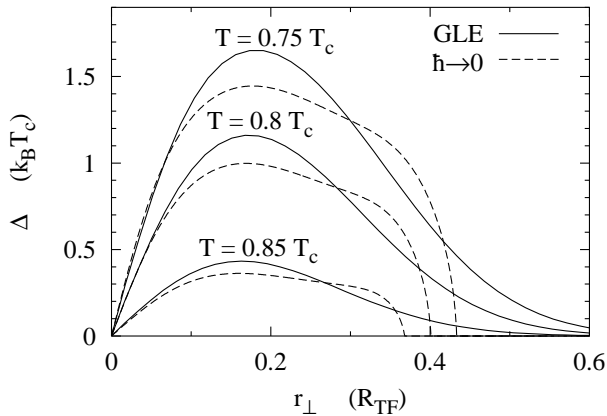,width=8.2cm}
\end{center}
\caption{\label{FigDelta} Order parameter $\Delta$ of a superfluid
trapped Fermi gas in a vortex state with $\kappa = 1$ as a function of
$\rperp$ for $z = 0$. The parameters were chosen corresponding to
573000 $^6$Li atoms in a spherical trap with $\omega = 2\pi\times 144$
Hz. The order parameter $\Delta$ is given in units of $k_B T_c$ ($T_c
= 31.2$ nK), the radius $\rperp$ in units of $R_{TF}$ ($R_{TF} = 48.7$
$\mu$m). Solid lines result from the numerical solution of the GLE
(\ref{gle}), whereas the dashed lines are obtained within the
($\hbar\to 0$) TF approach. The three pairs of curves correspond to
three different temperatures (from bottom to top: $0.85 T_c$, $0.8
T_c$, and $0.75 T_c$)}
\end{figure}

In Fig. \ref{FigDelta}
%
%%%%%%%%%%%%%%%%%%%%%%%%%%%%%%%%%%%%%%%%%%%%%%%%%%%%%%%%%%%%%%%%%%%%%%%%%
% Figure 6
%
%%%%%%%%%%%%%%%%%%%%%%%%%%%%%%%%%%%%%%%%%%%%%%%%%%%%%%%%%%%%%%%%%%%%%%%%%
%
we show the order parameter $\Delta$ obtained from the numerical
solution of Eq. (\ref{gle}) (solid lines). For the parameters listed
above, the lowest temperature for which vortex states can exist, i.e.,
the temperature for which $\tilde{\mu} = 5/2$, is approximately 0.86
$T_c$. Therefore we display the order parameter only for temperatures
below this value, namely for $T/T_c = 0.85$, 0.8, and 0.75. For $T/T_c
= 0.85$ the order parameter is still very small, and it increases
rapidly as the temperature decreases. As can be seen in Fig.
\ref{FigDelta}, in all three cases the ``amplitude'' of the order
parameter and the position of the maximum are well reproduced by our
TF approximation. Note that also the vortex core is well described.

As already stated, our TF solution has to be interpreted in terms of
distributions, and then the agreement is even better than it seems
from Fig. \ref{FigDelta} if one looks at integrated quantities. As
an example we consider the normalization $N = \int d^3 \tilde{r}
|\tilde{\phi}(\tilde{\vec{r}})|^2$. For the three temperatures
mentioned above, the normalizations obtained from the numerical
solution of the GLE and those corresponding to our TF approximation
are in good agreement, as shown in Table \ref{table2}.
%
%%%%%%%%%%%%%%%%%%%%%%%%%%%%%%%%%%%%%%%%%%%%%%%%%%%%%%%%%%%%%%%%%%%%
% Table II
%
\begin{table}
\begin{tabular}{llrl}
\hline\noalign{\smallskip}
$T/T_c$ & &\multicolumn{1}{c}{$N$} & \multicolumn{1}{c}{$F_{GL}$ ($\mu$K)} \\
\noalign{\smallskip}\hline\noalign{\smallskip}
  & GLE          &   2.12 &    -0.0079\\
\raisebox{1.5ex}[0pt]{$0.85$}
  & $\hbar\to 0$ &   2.37 &    -0.0088\\
\noalign{\smallskip}\hline\noalign{\smallskip}
  & GLE          &  16.37 &    -0.391 \\
\raisebox{1.5ex}[0pt]{$0.8$}
  & $\hbar\to 0$ &  18.41 &    -0.440 \\
\noalign{\smallskip}\hline\noalign{\smallskip}
  & GLE          &  34.18 &    -1.44  \\
\raisebox{1.5ex}[0pt]{$0.75$}
  & $\hbar\to 0$ &  38.20 &    -1.61  \\
\noalign{\smallskip}\hline
\end{tabular}
\caption{\label{table2} Normalization $N$ of the order parameter and GL
free energy $F_{GL}$ for the parameters used in Fig. \ref{FigDelta},
obtained from the numerical solution of the GLE and from our ($\hbar\to 0$)
TF approximation.}
\end{table}
%%%%%%%%%%%%%%%%%%%%%%%%%%%%%%%%%%%%%%%%%%%%%%%%%%%%%%%%%%%%%%%%%%%%%%%%
%
As a more meaningful example for an integrated quantity let us look at
the GL free energy $F_{GL}$. The explicit expression for the
functional $F_{GL}[\Delta]$ is given in Ref.
\cite{Baranov}. Following this reference, we retain only the leading
terms in the small quantities $K$, $\vec{r}/R_{TF}$, $\ln(T_c^{(0)}/T)$, and
$\Delta/(k_B T_c)$. Then, after integration by parts, the GL free energy
functional becomes
\begin{multline}
F_{GL}[\Delta]=\frac{mk_F}{2\pi^2\hbar^2}\int\! d^3r\,
  \Big\{-K^2 R_{TF}^2 \Delta^*\vec{\nabla}^2\Delta\\
  +\Big[\Big(\frac{1}{2\lambda}+1\Big)\frac{\vec{r}^2}{R_{TF}^2}
    -\ln\frac{T_c^{(0)}}{T}\Big]|\Delta|^2\\
  +\frac{7\zeta(3)}{16\pi^2}\,\frac{1}{(k_B T)^2}|\Delta|^4\Big\}\,.
\end{multline}
In the TF approach, the first term ($\propto \Delta^* \vec{\nabla}^2
\Delta$) cannot be obtained directly from the TF approximation for
$\Delta(\vec{r})$, but it rather has to be calculated analogous to the
kinetic energy density in Eq. (\ref{tau}). As a consequence, most of
the terms cancel, as it is the case if $\Delta(\vec{r})$ is the exact
solution of the GLE (\ref{gle}), and only the last term ($\propto
|\Delta|^4$) survives, but with negative sign. Thus, for the TF
approximation as well as for the exact solution of the GLE, we can
write in terms of the dimensionless variables defined above
\begin{equation}
F_{GL} = \frac{4 \epsilon_F^2(k_BT)^2}{\pi^2(\hbar\omega)^3}\,
  K^{5/2}\Big(\frac{2\lambda}{1+2\lambda}\Big)^{1/4}\int\! d^3\tilde{r}\,
  \Big(-\frac{\tilde{g}}{2}|\tilde{\Phi}|^4\Big)\,.
\end{equation}
Results for $F_{GL}$ obtained from the numerical solution of Eq.
(\ref{gle}) and from the TF solution are listed in Table
\ref{table2}. The agreement is as good as for the normalizations
$N$. In fact, the deviations are mainly due to the different
normalizations, i.e., the ratio $F_{GL}/N$ obtained from the TF
approximation is very close to the exact one.

To conclude this section, we stress that in the range of validity of
the GLE the ``chemical potential'' $\tilde{\mu}$ must not become
large. This is the reason for the rather small normalizations of the
order parameter and results in a shape of the order parameter as a
function of $\vec{r}$ which resembles very much the shape of a
non-interacting HO wave function. Under these conditions it is clear
that the $N\to\infty$ limit cannot be used as an approximate solution
of the GLE, as has also been noted in Ref. \cite{Rodriguez}.
%
%%%%%%%%%%%%%%%%%%%%%%%%%%%%%%%%%%%%%%%%%%%%%%%%%%%%%%%%%%%%%%%%%%%%%
\section{Summary and conclusions}
Using the semiclassical Thomas-Fermi approximation understood as
$\hbar\to 0$ limit rather than $N\to\infty$ limit, we have studied the
vortex states of a Bose condensate of atoms confined in a spherical
magnetic trap.

We started analyzing the vortex states in a non-interacting trapped
Bose gas. Due to the symmetry of the problem, we have obtained first
the Thomas-Fermi density projected on states of defined $L_z$. In this
non-interacting case the density is normalized by adjusting the
normalization constant $c_\kappa$, and the chemical potential $\mu$ is
fixed, according to the WKB quantization rule, to the quantal
eigenvalue of the quantum state.

In the interacting case the normalization constant and the chemical
potential are fixed to normalize the Thomas-Fermi density to the
number of particles and that the integrated level density become equal
to that of the non-interacting case. For particle numbers where the
kinetic energy coming from the radial and axial motion is a
non-negligible part of the total kinetic energy, our Thomas-Fermi
approach, understood as $\hbar\to 0$ limit, yields very satisfying
results as compared with the corresponding quantal values. For a very
large number of particles in the condensate, the small Thomas-Fermi
kinetic energy, obtained in our approach, is smaller than the quantal
kinetic energy, which, for large number of particles, is also
dominated by quantal corrections as it happens for the ground state
\cite{SchuckVinas}.

The vortex state density profiles obtained in our Thomas-Fermi
approximation reproduce quite well the quantal ones, especially for a
very large number of particles. However, inside the vortex core our
Thomas-Fermi densities are too high. Also near the classical turning
point our TF densities locally fail because at this point the density
is completely dominated by quantal contributions which are
non-analytical in $\hbar$ and which cannot be reproduced by
semiclassical approximations of the TF type. However, it shall be kept
in mind that the semiclassical density has to be understood as
distribution very efficient for describing expectation values rather
than local quantities such as the density profile. In this sense we
see that the quantities presented in Table \ref{table1} are much more
accurate than one would expect from an inspection of the local
densities shown in Figs. \ref{FigphiRb2}-\ref{FigphiRb6}.

The approach is also well suited for the description of vortex states
of superfluid trapped fermionic systems in the GL regime, where the
various approximations developed for large $N$ cannot be used at
all. It should be mentioned that the conditions for the validity of
the GLE imply that the parameters of the equivalent GPE always
correspond to a rather small number of particles. In this case the
normalization of the order parameter (see Table \ref{table2}) and the
position of the maximum are well reproduced by the $\hbar\to 0$ limit
as compared with numerical solutions of the GLE. Also the vortex-core
region is well described in this case.
%
%%%%%%%%%%%%%%%%%%%%%%%%%%%%%%%%%%%%%%%%%%%%%%%%%%%%%%%%%%%%%%%%%%%%%%%%
%
\begin{acknowledgments}
We are indebted to A. Polls for supplying us the GP code and to him,
M. Guilleumas, and S. Stringari for useful comments. One of us (M.U.)
acknowledges support by the Alexander von Humboldt foundation
(Germany) as a Feodor-Lynen fellow. X.V. acknowledges financial
support from DGCYT (Spain) under grant Pb98-1247, from DGR (Catalonia)
under grant 2000SGR-00024, and from the CICYT-IN2P3 collaboration.
\end{acknowledgments}
%%%%%%%%%%%%%%%%%%%%%%%%%%%%%%%%%%%%%%%%%%%%%%%%%%%%%%%%%%%%%%%%%%%%%%%%
%
\appendix*
\section{Large-$N$ limit for vortex states}
In order to discuss the large-$N$ limit for vortex states more
thoroughly, we start from Eq. (\ref{philocal}), which, as shown in
Fig. \ref{FigphiRb6}, becomes very accurate in the limit of large
$N$. Let us look at the different contributions to the kinetic energy,
$\erot$ and $\ekin$. In the infinite system, the energies per unit
length, $dE_{\mathit{rot}}/dz$ and $dE_{\mathit{kin}}/dz$ can easily
be obtained from the numerical solution for $f_\kappa$. Since
$dE_{\mathit{rot}}/dz$ diverges logarithmically, the corresponding
integral has to be cut off at some radius $R$ \cite{Ginzburg}. For
$\kappa = 1$ the results read ($R\gg\xi_0$):
\begin{align}
\frac{dE_{\mathit{rot}}}{dz}
  &= \frac{\pi\hbar^2\rho_0}{m}\Big(\ln \frac{R}{\xi_0}-0.40\Big)\,,\\
\frac{dE_{\mathit{kin}}}{dz}
  &= 0.28 \frac{\pi\hbar^2\rho_0}{m}\,.
\end{align}
In complete analogy to the derivation of the total energy of a vortex
in a trapped system in Ref. \cite{Lundh}, one can use these results
to obtain explicit expressions for the rotational and radial kinetic
energies of a vortex, $\erot$ and $\ekin^{\mathit{core}}$, which for a
spherical trapping potential read
\begin{align}
\erot &= \frac{1}{N}\,\frac{4\pi\rho_0}{3}\,\frac{\hbar^2}{m}\,
  r_{\mathit{max}}\,\Big(\ln\frac{r_{\mathit{max}}}{\xi_0}-1.18\Big)\,,\\
\ekin^{\mathit{core}} &= 0.28\,\frac{1}{N}\,\frac{4\pi\rho_0}{3}\,
  \frac{\hbar^2}{m}\,r_{\mathit{max}}\,.
\end{align}
Here $r_{\mathit{max}} = \sqrt{2\mu/(m\omega^2)}$ is the radius of the
condensate. However, the kinetic energy has also another contribution
$\ekin^{\mathit{trap}}$ due to the finite size of the trapped
system. Since outside the vortex core the shape of the condensate is
almost not changed, we assume that for this contribution the relation
derived in Ref. \cite{Dalfovo96c} for the case without vortex,
remains valid:
\begin{equation}
\ekin^{\mathit{trap}} = \frac{5}{2}\,\frac{\hbar^2}{m r_{\mathit{max}}^2}
  \Big(\ln\frac{r_{\mathit{max}}}{\aHO}-0.26\Big)\,.
\label{ekindalfovo}
\end{equation}
Since the volume of the vortex core is negligible in the limit
$N\to\infty$, $\mu$ depends on $N$ in the same way as in the large-$N$
limit for the ground state,
\begin{equation}
\mu = \frac{\hbar\omega}{2} \Big(\frac{15 N a}{\aHO}\Big)^{2/5}\,.
\end{equation}
Using this, we finally obtain
\begin{align}
\erot &= \hbar\omega \Big(\frac{\aHO}{15 N a}\Big)^{2/5}
  \Big(\ln\frac{15 N a}{\aHO}-2.95\Big)\,,\label{erotlargen}\\
\ekin &= \hbar\omega \Big(\frac{\aHO}{15 N a}\Big)^{2/5}
  \Big(\frac{1}{2}\ln\frac{15 N a}{\aHO}-0.51\Big)\,.\label{ekinlargen}
\end{align}

From these equations we conclude that the ratio $\ekin/\erot$ does not
go to zero, but approaches $1/2$ for $N\to\infty$. Hence, neglecting
the radial and axial parts of the kinetic energy, but retaining the
rotational part, as it is done in the literature
\cite{Fetter,Rokhsar,Ho,Sinha}, is not justified and does not
correspond to the proper $N\to\infty$ limit. Instead, the correct
large-$N$ limit is given by Eq. (\ref{philocal}), except at the
surface of the condensate. The latter can be approximated, e.g., by
the exact solution of the GPE for a linear potential, as it has been
done in Ref. \cite{Dalfovo96c} in order to derive Eq.
(\ref{ekindalfovo}), and also in Ref. \cite{Lundh}.

It should, however, be noted that Eq. (\ref{philocal}) and Eq.
(\ref{ekindalfovo}) correspond to a partial resummation to all orders
in $\hbar$ as demonstrates the nonanalytical dependence on $\hbar$ of
these quantities. Such resummation techniques, also encountered in the
WKB approximation, are necessary whenever the asymptotic
Wigner-Kirkwood $\hbar$ expansion breaks down. This is always the case
when the gradients of the potential start to diverge like in the
vortex core for $N\to\infty$, see Fig. \ref{FigVsc}.
%
%%%%%%%%%%%%%%%%%%%%%%%%%%%%%%%%%%%%%%%%%%%%%%%%%%%%%%%%%%%%%%%%%%%%%%%%***    
%

\end{document}